\documentclass{article}

\usepackage{arxiv}
\usepackage{float}
\usepackage[caption = false]{subfig}
\usepackage{graphicx}

\usepackage[utf8]{inputenc} 
\usepackage[T1]{fontenc}    
\usepackage{hyperref}       
\usepackage{url}            
\usepackage{booktabs}       
\usepackage{amsfonts}       
\usepackage{amsmath}
\usepackage{cleveref}
\usepackage{nicefrac}       
\usepackage{microtype}      
\usepackage{lipsum}
\usepackage{xcolor}

\title{Hierarchical Bayesian propulsion power models for marine vessels}

\author{
  Antti Solonen\\
  School of Engineering Science\\
  LUT University\\
  \texttt{antti.solonen@gmail.com}\\ 
  \and
  Ramona Maraia\\
  School of Engineering Science\\
  LUT University\\
  \texttt{ramona.maraia@lut.fi} \\
  \and
  Sebastian Springer\\
  Research unit of Mathematical Sciences\\
  University of Oulu\\
  \texttt{sebastian.springer@oulu.fi}\\
   \and
  Heikki Haario\\
  School of Engineering Science\\
  LUT University\\
  Earth Observation  Research\\
  Finnish Meteorological Institute\\
  \texttt{heikki.haario@lut.fi}\\
  \and
  Marko Laine\\
  Meteorological Research\\
  Finnish Meteorological Institute\\
  \texttt{marko.laine@fmi.fi}\\
  \and
  Olle Räty\\
  Meteorological Research\\
  Finnish Meteorological Institute\\
  \texttt{olle.raty@fmi.fi}
  \and
  Jukka-Pekka Jalkanen\\
  Atmospheric Composition Research\\
  Finnish Meteorological Institute\\
  \texttt{jukka-pekka.jalkanen@fmi.fi}\\
  \and
  Matti Antola\\
  Eniram, a Wärtsilä Company\\
  \texttt{matti.antola@wartsila.com}\\
 
}
\graphicspath{{Figures/}} 
\begin{document}
\maketitle

\begin{abstract}
Assessing the magnitude of fuel consumption of marine traffic is a challenging task. The consumption can be reduced  by the ways the vessels are operated, to achieve both improved cost efficiency and reduced CO2 emissions. Mathematical models for predicting ships' consumption are in a central role in both of these tasks. Nowadays, many ships are equipped with data collection systems, which enable data-based calibration of the consumption models. Typically this calibration procedure is carried out independently for each particular ship, using only data collected from the ship in question. In this paper, we demonstrate a hierarchical Bayesian modeling approach, where we fit a single model over many vessels, with the assumption that the parameters of vessels of same type and similar characteristics (e.g. vessel size) are likely close to each other. The benefits of such an approach are two-fold; 1) we can borrow information about parameters that are not well informed by the vessel-specific data using data from similar ships, and 2) we can use the final hierarchical model to predict the behavior of a vessel from which we don't have any data, based only on its characteristics. In this paper, we discuss the basic concept and present a first simple version of the model. We apply the Stan statistical modeling tool for the model fitting and use real data from 64 cruise ships collected via the widely used commercial Eniram platform. By using Bayesian statistical methods we obtain uncertainties for the model predictions, too. The prediction accuracy of the model is compared to an existing data-free modeling approach.
\end{abstract}


\section{Introduction}

Marine vessels are a large contributor to global CO2 emissions\footnote{Global annual CO2 emissions due to shipping were estimated to be 938 million tonnes in 2012 \cite{imo-ghg3} and 831 million tonnes in 2015 \cite{johansson2017}. A single large ship can burn 40000 tons of fuel and produce 120000 tons of CO2 per year.}. Lately, emphasis has been put on optimizing various aspects of vessel operations, such as route and speed profile selection, which helps in reducing the emissions and make shipping more cost efficient. To be able to run such optimization, predictive models of vessels' fuel consumption are needed.

Moreover, vessel consumption models can be used to assess the global emissions of shipping. One such \textit{bottom-up} approach, where a consumption model is built for essentially every major ship in the world, is described in \cite{jalkanen09,jalkanen12, johansson2017}. The approach utilizes existing methods for ship resistance calculations, where various resistance coefficients are estimated based on different ship characteristics that can be obtained from commercial ship databases\footnote{for instance IHS Markit \url{https://ihsmarkit.com/}}. Obtained models are simulated using vessels' AIS data\footnote{AIS (Automatic Identification System) is a system through which vessels report their location and speed. The International Maritime Organization (IMO) requires that AIS is used in all ships with gross tonnage larger than 300.}. Another model-based approach for assessing emissions is described in \cite{smith2012}. Both of these approaches utilize \textit{white box} modelling, which means that vessel consumption data are not used in training the models. Including such data into the modelling (a \textit{grey box} approach) will improve the accuracy of the models. Moreover, the white box modelling usually neglects some major resistance factors such as wind, waves, shallow water resistance and hull fouling, which can influence the vessels' total resistance and contribute significantly to consumption.

Nowadays vessel-specific operational data related to vessel's consumption are becoming increasingly available. With such data, accurate models can be calibrated for each vessel. See, e.g., the work in \cite{coraddu15,coraddu17} for comparison between the white, grey and black box modeling approaches, and a comprehensive pipeline for building data-based vessel consumption models. Combining data and physics in the modeling enables detailed optimization of vessel operations, monitoring the vessel's propulsion performance and other detailed ship-specific analytics. Various companies offer such solutions, including Eniram Ltd, a Wärtsilä Company, the collaboration partner in this study\footnote{Part of Wärtsilä, see \url{https://www.wartsila.com/eniram}}. Due to the reasons listed above the grey box approach has been selected for propulsion power modelling at Eniram. 
 
However, collecting detailed high fidelity data is costly, which calls for methods to build models also for ships for which we have limited or no data available. For instance, high-fidelity data based on high-frequency logging onboard a ship might be available only for a small amount of ships, but there might be, e.g., noon-report type of consumption data available for a larger number of ships, where the crew has reported total consumption numbers over certain time intervals (e.g. 24h). Such data is being collected in increasing amounts due to EU MRV and IMO DCS regulations that require consumption reporting for vessels with gross tonnage (GT) higher than 5000. Calibrating consumption models with noon-report data is challenging and calls for statistical methods to include all available information into the resistance coefficient estimation. See \cite{antola17-noon} for some discussion on dealing with such scarce data.

The goal of this paper is to illustrate an approach where we can use the data collected from a group of ships, and generalize the information to a larger population of vessels. We use real data collected from 64 cruise ships via the Eniram platform, anonymized due to data ownership questions. Our approach is to build a hierarchical Bayesian model that encompasses all the vessel-specific parameters and coefficients, but also includes a "hyper-model" that links the coefficient values between ships together. The approach is based on the idea that the resistance coefficients between two ships of similar characteristics (e.g. type and dimension) are likely close to each other. Both the vessel specific coefficients and the hyper-model parameters defining between-ship relationships are learned from the available data. The novelty compared to the existing resistance calculations is that the consumption model parameter values are informed by the data, and can thus give more accurate predictions than the classical methods. Moreover, estimating the resistance coefficients for a ship that has only limited data available can be made more robust and stable by including information about other similar ships. For instance, using only a small amount of noon-report consumption data can lead to nonphysical resistance coefficient estimates, but including the hyper-model can help significantly, as demonstrated later in this paper. Finally, the "hyper-model" can be used to predict the consumption of a ship from which we have no data, based only on its characteristics, which enables applications such as the global emission estimation discussed above and optimization of ships operations (e.g. route, speed) at scale, without involving expensive data collection platforms on-board.

To the best of our knowledge, using such hierarchical modeling setup in the context of marine vessel consumption modeling has not been discussed before. One straightforward existing approach for borrowing information between ships is to apply models built for a vessel to its sister vessels, see, e.g., \cite{bazari07,Bialystocki16} for discussion. Our approach generalizes this idea. With hierarchical modeling it is possible to learn from the data how the ship coefficients link together via various ship characteristics. Thus, one can "interpolate" between existing ship models and does not need to have an exact sister vessel model available.

We present a prototype of the hierarchical model and show that even such simple data driven approach can compete in prediction accuracy with the classical resistance calculations. We demonstrate how the regularization effect of the hierarchical model makes the results more stable and robust compared to independent vessel-specific models. Due to simplicity and data availability, we restrict ourselves to cruise ships and propulsion power modelling. Here, the goal is to present the hierarchical modeling concept with simple examples; more work is required to increase the sophistication of the model formulations, to generalize to other ship types and to include service power models and engine models to turn power consumption into fuel consumption. Moreover, to assess the CO2 emissions of ships, one needs to have models for mapping fuel consumption into emissions, see \cite{jalkanen09,jalkanen12,johansson2017} for discussion on these topics. Although we focus here only on propulsion power modeling, similar hierarchical modeling ideas are likely applicable to service power and engine modeling as well.

The paper is organized as follows. Section~\ref{sec:general} describes the general setup and the applied models. Section~\ref{sec:results} describes the numerical examples and results. Section~\ref{sec:conclusions} concludes the paper.

\section{Modeling Setup}
\label{sec:general}

This section gives an overview of the two approaches used to model propulsion power consumption $P$, which is the target variable in this paper. We first briefly present an existing White Box modeling approach developed in \cite{jalkanen12} that will be used in the numerical comparisons in Section~\ref{sec:results}. We then introduce the data-driven, hierarchical Grey Box model and the rationale behind it.

Note that the hierarchical modeling approach targets the case where we have no consumption data available for a ship, only a (possibly limited) set of vessel characteristics. That is, our aim is not to build the most sophisticated grey box model based on high-fidelity vessel-specific data, but to demonstrate the hierarchical concept using very simple grey box models that consider only frictional and wind resistance. For building grey box models using high-fidelity data, we refer the reader to \cite{coraddu15, coraddu17}, and to the work of some commercial players in the field, see \cite{solonen16, haranen16, antola17-stw}. Extending the hierarchical modeling concept to more sophisticated vessel models that take more resistance effects into account is left for future work.

\subsection{White Box approach: STEAM2}
\label{sec:STEAM2}

As a comparison to the hierarchical model presented in the next Section, we use the STEAM2 model developed in \cite{jalkanen12}, which builds on earlier work, such as the widely used Hollenbach resistance calculations, see, e.g., \cite{schneekluth98, hollenbach98}. The purpose of STEAM2 is to build a white box model for basically every major ship in the world using only publicly available ship characteristics, with the goal of assessing the emissions from shipping. To our knowledge, STEAM2, together with the simpler modeling approach described in \cite{smith2012}, is one of the only systems capable of doing such bottom-up global simulations of consumption and emissions.

In STEAM2, the propulsion power consumption is calculated simply via $P=R_TV$, where $V$ is vessel speed through water and $R_T$ is the total resistance. In this approach, the total resistance is approximated by
\begin{equation}
    R_T = R_F + R_R,
\end{equation}
where $R_F$ is the frictional resistance between the water and the vessel's wet surface, and $R_R$ is the "residual resistance" that accounts for other hydrodynamic resistance components such as wave making (the power needed for forming the wave pattern that the vessel generates).

The frictional resistance is calculated as
\begin{equation}
    R_F = C_F\frac{\rho}{2} S V^2,
\end{equation}
where $S$ is vessel's wet surface area, $\rho$ is water density and $C_F$ is the frictional resistance coefficient. Here, we follow the widely used ITTC approach\footnote{\url{https://ittc.info/}}, where $C_F=0.075/(\log_{10}(R_n) -2)$ and $R_n$ is the Reynolds number, calculated here as $R_n=V L_{wl}/\nu$, where $L_{wl}$ is the waterline length of the vessel and $\nu$ is the kinematic viscosity. 

The waterline length $L_{wl}$ is typically available from various commercial ship databases such as IHS Markit, but the wet surface area $S$ is typically unknown. In the Hollenbach calculations, $S$ is estimated using a rather complicated formula that involves various ship dimensions. The formula was obtained in \cite{hollenbach98} via regression analysis applied to model tank test results for 433 ships. The formula and the regression coefficients are reported, for instance, in \cite{schneekluth98} and are not reproduced here for brevity.

The residual resistance is calculated as 
\begin{equation}
    R_R = C_R\frac{\rho}{2} \left( \frac{B \cdot T}{10} \right)  V^2,
\end{equation}
where $B$ is vessel breadth, $T$ vessel draft and $C_R$ the residual resistance coefficient. The residual resistance coefficient $C_R$ is obtained via the Hollenbach method using a similar approach than for wet surface area; the formula and the best-fit regression coefficients are reported in \cite{schneekluth98} and are not reproduced here.

Note that in our simple data-based model described in \cref{sec:simple}, we clump everything in front of $V^2$ into one constant parameter. We thus ignore the fact that the frictional resistance reduces a bit as a function of vessel speed. However, this effect is small compared to the overall accuracy of the models, and our goal is to show that a simple parameterization can give prediction accuracy comparable to more complex white box formulations. Moreover, the data-based approach allows us to fine tune the coefficients for each vessel instead of using fixed formulas and coefficients. 

 In addition to the Hollenbach resistance formulas behind STEAM2, there are other white box resistance calculation approaches that could be used to generate ship models based on vessel characteristics. For instance, the Holtrop-Mennen method is a similar, widely used approach. See \cite{birk19} for discussion and comparison between the methods. It would be interesting to compare the hierarchical method to Holtrop-Mennen as well, but it requires a more extensive set of ship characteristics that often is not easily available.

Note that many resistance components are ignored in STEAM2 (and in many other white box methods like the mentioned Holtrop-Mennen), such as the aerodynamic resistance that we include in the data-based model. Other ignored resistance effects include, for instance, wave breaking resistance (resistance caused by the waves that the vessel needs to propel through), shallow water resistance (additional resistance caused by sailing in shallow waters, also known as squatting, see \cite{schneekluth98} for more details) and hull roughness and biofouling. These effects could be included in the resistance calculations, but then we would need to come up with values for the corresponding resistance coefficients, which can be complex and require many simplifying assumptions and new vessel characteristics (at present, the STEAM2 model is being developed to this direction). This is in contrast with grey box models -- and thus also the hierarchical grey box approach described in the following Section -- where it's possible to learn the required coefficients from data and these extra effects could be added in a rather straight forward manner.

\subsection{Simplified hierarchical Grey Box model}
\label{sec:simple}

Here we present a simple prototype for the hierarchical Bayesian propulsion power model, which is the main novelty of this paper. Let us assume that we have a ship-specific propulsion power model like
\begin{equation}
    P_i = f(x_i, \theta_i)+\varepsilon, \quad i=1,..., N,
\end{equation}
where $i$ is the ship index, $P_i$ is the observed propulsion power, $x_i$ are the observed model inputs (vessel speed, wind speed and angle, etc.), and $\theta_i$ are the unknown parameters that we want to estimate (e.g. various resistance coefficients). The error term $\varepsilon_i$ denotes the discrepancy between modelled and observed propulsion power.

\begin{figure}
  \centering
  \includegraphics[width=0.9\linewidth]{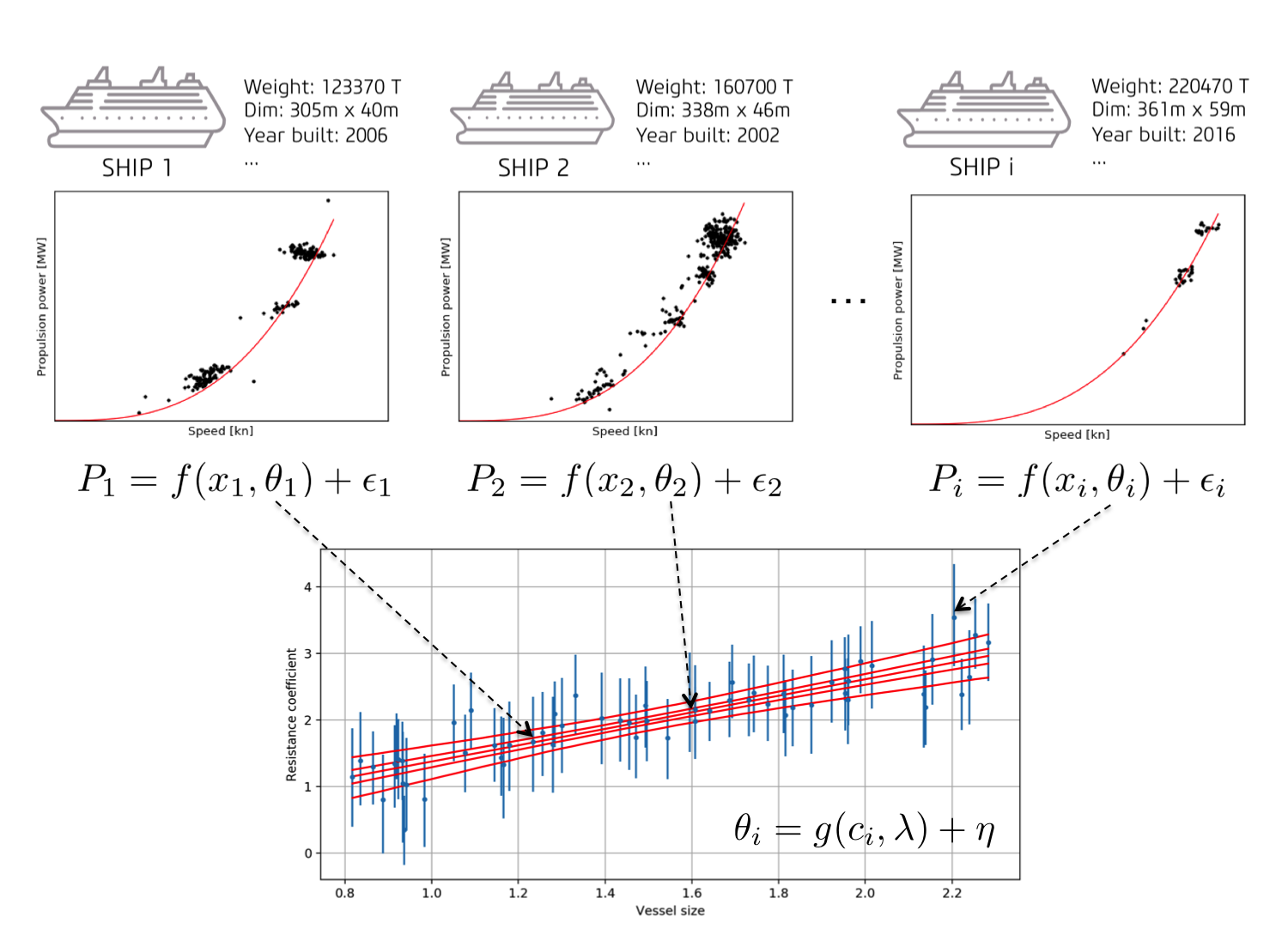}
  \caption{Illustration of the hierarchical model. Based on ship-specific power consumption data and various model inputs (black dots) the goal is to learn both ship-specific parameter values (blue error-bars) and hyper-parameters that link the between-ship parameters together. The ships in the figure are imaginary; the graph is generated with the synthetic demo code available in \url{https://github.com/solbes/stanship}.}
  \label{fig:hier_demo}
\end{figure}

The traditional approach would be to estimate the parameters for each ship independently, using the data $(x_i, P_i)$ for each ship. Here, instead, we add another layer of modeling; we assume that the parameter values can be predicted with some (unknown) accuracy using various ship characteristics. Thus, we write a model for the ship-specific parameters as
\begin{equation}
    \theta_i = g(c_i, \lambda)+\eta,
\end{equation}
where $c_i$ denotes the characteristics, $\lambda$ is a vector of unknown hyper-parameters and $\eta$ describes how accurate this hyper-model $g$ is in predicting the parameter values. The ship characteristics could be related, for instance, to vessel's size (e.g. weight), dimensions (width and length), construction year, or any other vessel metadata that carries some information about $\theta_i$. The general setup is illustrated in Figure~\ref{fig:hier_demo}.

The goal is now to learn both the ship-specific coefficients $\theta_i$ and the hyper-parameters $\lambda$ using all the observed data $P_{1:N}$. In Bayesian terms, this amounts to finding the posterior distribution of the parameters given the measured data, $p(\theta_{1:N}, \lambda | P_{1:N})$. In addition, we would like to learn about the error terms $\varepsilon$ and $\eta$, which can be done by fixing the form of the error distributions (e.g. zero mean Gaussians) and including the parameters of the error distributions (e.g. variances of the Gaussians) to the group of parameters that are estimated from the data. 

Finally, when we have learned the posterior distribution for all the parameters, we have a model where the ship-specific coefficients are informed by both their own data and data from a similar ship. Full Bayesian analysis of the parameters also lets us predict the behavior of a vessel that is not included in the training data, and give an idea about how certain we are about the predicted behavior. This feature is missing from the classical resistance calculations.

We demonstrate the hierarchical modeling idea with a simple example. Our vessel-specific model includes only two terms; one describing hydrodynamic resistances (e.g. friction and wave making) and one for aerodynamic resistance. Propulsion power for the ship $i$ is calculated via $P_i=R_{T,i}V_i$, where $R_{T,i}$ is the total resistance, which is here approximated as the sum of hydrodynamic and aerodynamic resistances:
\begin{equation}
    R_{T,i} = R_{H,i} + R_{A,i}.
\end{equation}

The hydrodynamic resistance model used here is quite crude; we simply state that the resistance increases proportionally to vessel speed squared: $R_{H,i}=a_iV_i^2$, where $a_i$ is the hydrodynamic resistance coefficient, which is assumed to be an unknown constant. In reality, the hydrodynamic resistance coefficient is not constant though; it varies as a function of vessel speed and draft, for instance. However, for demonstration purposes this approximation is adequate, especially for cruise ships considered in this study for which the draft variations are minimal.

For wind, we use the simple approximation that the wind resistance is proportional to relative wind speed squared. When we project the wind resistance force vector to the heading of the ship, we get $R_{A,i} = b_i\cos(\alpha_i)U_{R,i}^2$, where $\alpha_i$ is the relative wind angle, $U_{R,i}$ is the relative wind velocity and $b_i$ is the unknown wind resistance coefficient. Note that this approximation is rather crude; it assumes, for instance, that the contact area between the wind and the vessel hull is constant. Some more sophisticated wind formulas, such as those described in \cite{blendermann1996, schneekluth98}, could be taken into use, but this simple formula is sufficient for demonstration purposes again.

With these approximations, our simplified propulsion power model for ship $i$ reads as
\begin{equation}
    P_{i} = a_iV_i^3 + b_i\cos(\alpha_i)U_{R,i}^2V_i + \varepsilon_i.
\end{equation}
Now, the goal is to estimate coefficients $a_i$ and $b_i$ from measured data. This could be done individually for each ship, but that could be problematic if the data is not very informative about the coefficients. That is why we include the hyper-model to tie the coefficients between ships together in one model.

The task of the hyper-model is to predict the values of the resistance coefficients based on some ship characteristics $c_i$. Here, we use the ships total weight $w_i$ (gross tonnage, GT) as the hyper-model input, and model both coefficients as linear functions of GT:
\begin{equation}\label{eq:Aeb}
  \begin{split}
     a_i &= \lambda_1 + \lambda_2w_i  + \eta_a \\
     b_i &= \lambda_3 + \lambda_4w_i  + \eta_b,
  \end{split}
\end{equation}
where $\eta_a$ and $\eta_b$ are Gaussian error terms.

This is obviously not a very physical model. In more realistic settings, one could model the hydrodynamic and aerodynamic resistance using the vessel's dimensions, for instance. Here we pick GT as the input variable since it is easily available for all ships. A better indicator of vessel's mass might be its deadweight, or one could utilize the ship's block coefficient in some way, but here we use just GT based on easy data availability. The linear model choice comes from empirical observations; individual coefficients seem to roughly scale linearly as a function of vessel's GT. Note also that the ability to use such nonphysical parameterizations can be considered as a strength of the data-based approach; we can essentially insert any parameterization and try to use data to figure out the relationships between unknown coefficients and ship characteristics.

The remaining task is to estimate all of the ship-specific resistance coefficients in one model together with the hyper-model parameters $\lambda_i$. Moreover, as the vessel-specific model and hyper-model errors, $\varepsilon_i$, $\eta_a$ and $\eta_b$ are unknown, we will estimate them from the data, as well. We will assume that the errors are normally distributed and zero mean: $\varepsilon_i \sim N(0,\sigma_{i})$, $\eta_a \sim N(0,\sigma_a)$ and $\eta_b \sim N(0,\sigma_b)$. In addition to the resistance coefficients and hyper-model slopes and intercepts, we also estimate the variances $(\sigma_i, \sigma_a, \sigma_b)$. For Bayesian statistical analysis we need to specify prior uncertainties for all the model parameters. We use uniform priors for the resistance coefficients, and uniform priors with positivity constraints for the variance parameters. With less informative data or a smaller number of groups (ships), one might need to constrain the variance parameter more. See \cite{gelman2006} about setting priors for variance parameters in hierarchical models\footnote{See also \url{https://github.com/stan-dev/stan/wiki/Prior-Choice-Recommendations} about priors}.

The equations for the ship-specific models are simple and linear in parameters, but fitting the full hierarchical model is far from trivial. With 50+ ships the number of estimated parameters becomes rather high -- a few hundred -- and exploring this high-dimensional posterior distribution calls for efficient numerical methods. In recent years, flexible and openly available tools for defining and fitting such hierarchical Bayesian models have been developed, including, for instance, PyMC3 and the probabilistic programming language Stan \cite{pymc3, stan}. Here, the model fitting is carried out with the latter one, which implements a carefully tuned Markov Chain Monte Carlo (MCMC) sampler that is capable of exploring high-dimensional distributions. Model implementation with synthetic data (real data cannot be distributed) is available online\footnote{\url{https://github.com/solbes/stanship}}. The reader is referred to the experimental section for more details.

\section{Results}
\label{sec:results}

In this Section we present three numerical examples. The first illustrates how the hierarchical modelling regularizes the ship-specific parameter estimation. The second example compares the data-based model to the white box approach of Section~\ref{sec:STEAM2}. The last example demonstrates the ability to obtain uncertainty statistics for the model predictions.

\subsection{Description of the dataset}

We use real data in the experiments obtained from the Eniram platform to calibrate the grey box models. Propulsion power measurements are obtained from the vessels' automation systems. For vessel speed, we use speed over ground obtained from the vessel (a GPS-based measurement) augmented with ocean current forecasts to get an estimate of speed through water. For wind angle and wind speed, we use values from a weather forecast provider. Results and data are anonymized.

The original dataset contained very high-frequency data (time step was in the order of seconds). However, this is an overkill for the type of modeling we discuss in this paper, so the data was initially down-sampled to contain one value every 30 minutes. For some of the experiments below, we further averaged the data to daily resolution in order to emulate noon-report data.

Before calibrating the models, some data preprocessing was applied. First of all, a legging algorithm was used to take only data when the ship is in steady operation at sea (not at port or doing manouvering near the port). The details of this algorithm are propietary information. Moreover, we filtered out points when the depth of the ocean according to the GEBGO-2014 database was very low, in order to remove some extreme squatting (shallow water resistance) periods. Moreover, some values that were deemed erroneous (e.g. negative speed or power, unphysically high wind readings, etc.) were removed.

In the demo code package published in \url{https://github.com/solbes/stanship}, we generate a synthetic dataset that resembles the original dataset. However, in the experiments described below, the real dataset was applied.

\subsection{Regularizing effect of hierarchy}
\label{sec:regularization}

Here, we demonstrate how the hierarchical modeling can help to identify the parameters of individual ships, in the case where the ship-specific data are not informative about the unknowns.
We make the following experiment. Instead of modelling the momentary power consumption, we attempt to emulate a setting where we only have "noon-report" type of data available; that is, we have total consumption readings over given time intervals (e.g. 24~h) and momentary vessel speed and weather data with higher resolution. We choose this setting for demonstration purposes, since such aggregated data has obviously much less information about the parameters than the momentary data, and using noon-report data to calibrate ship models is thus challenging.

We model the total consumption over a given time interval by integrating both sides of the ship-specific power models. For ship $i$ and a single 24~h period the model now reads as
\begin{equation}
    \int_{24h}P_i(t)dt = a_i\int_{24h} V_i(t)^3dt + b_i\int_{24h} \cos(\alpha_i(t))U_{R,i}(t)^2V_i(t)dt + e_i,
\end{equation}
and thus the model remains linear with respect to the parameters. The training data for the ship-specific models are now the daily total consumption and integrated model input terms. In the experiment here, we replace the integrals by 24~h averages; in this way we don't need to worry if a few data points are missing from some 24~h intervals.

Note that in practice the noon-reported consumption is not necessarily reported at even 24h intervals. In real applications there is also added complexity related to handling maneuvering periods (when consumption can be unpredictable), missing data and other consumers in addition to propulsion (service power), for instance.

\begin{figure}[htb!]
  \centering
  \includegraphics[width=0.9\linewidth]{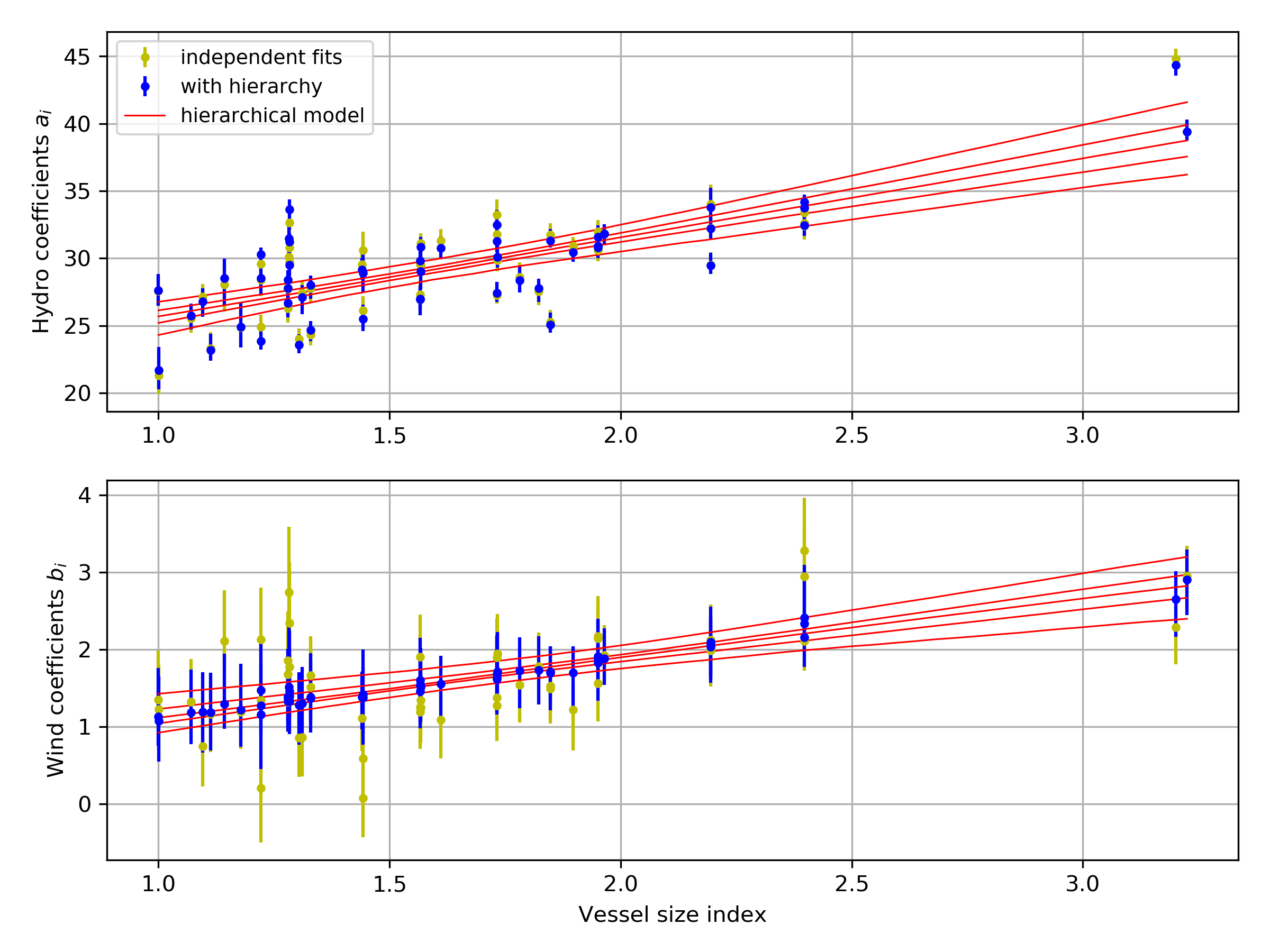}
  \caption{Illustration of the hierarchical model. Vessel size index is defined so that the smallest ship has value 1.}
  \label{fig:reg_demo}
\end{figure}

We use around 100 daily averages in the model fitting for each ship. The results for the hydrodynamic and aerodynamic coefficients for each ship with and without the hierarchy are given in \cref{fig:reg_demo}. We see that the hydrodynamic coefficients are well identified with the ship-specific data alone, and the hierarchy does not have much of an effect. However, for the wind resistance coefficients the situation is different. There is not enough information in the noisy data to calibrate the coefficients, and thus fitting ships independently yields some unrealistic values (e.g.\ close to zero) and the uncertainty is large. Adding the hierarchy pools the estimates closer to the linear prior and yields more reasonable looking estimates. We expect similar results for other resistance factors that might not be well-informed by the vessel-specific data, such as the shallow water resistance effect.

\subsection{Comparing the models}

In this Section the goal is to compare the presented hierarchical Bayesian grey box approach to the STEAM2 white box model discussed in \cref{sec:STEAM2}. We compare STEAM2 to two data-based models; one where the resistance coefficients are predicted using the prior model (vessel's gross tonnage), and one that uses the vessel-specific resistance coefficients obtained from the hierarchical model fit. Note that the latter would obviously not be available in cases where vessel consumption data is unavailable, but the results are presented here anyway for reference. 

\begin{figure}
    \centering
    \includegraphics[width = 5.7in]{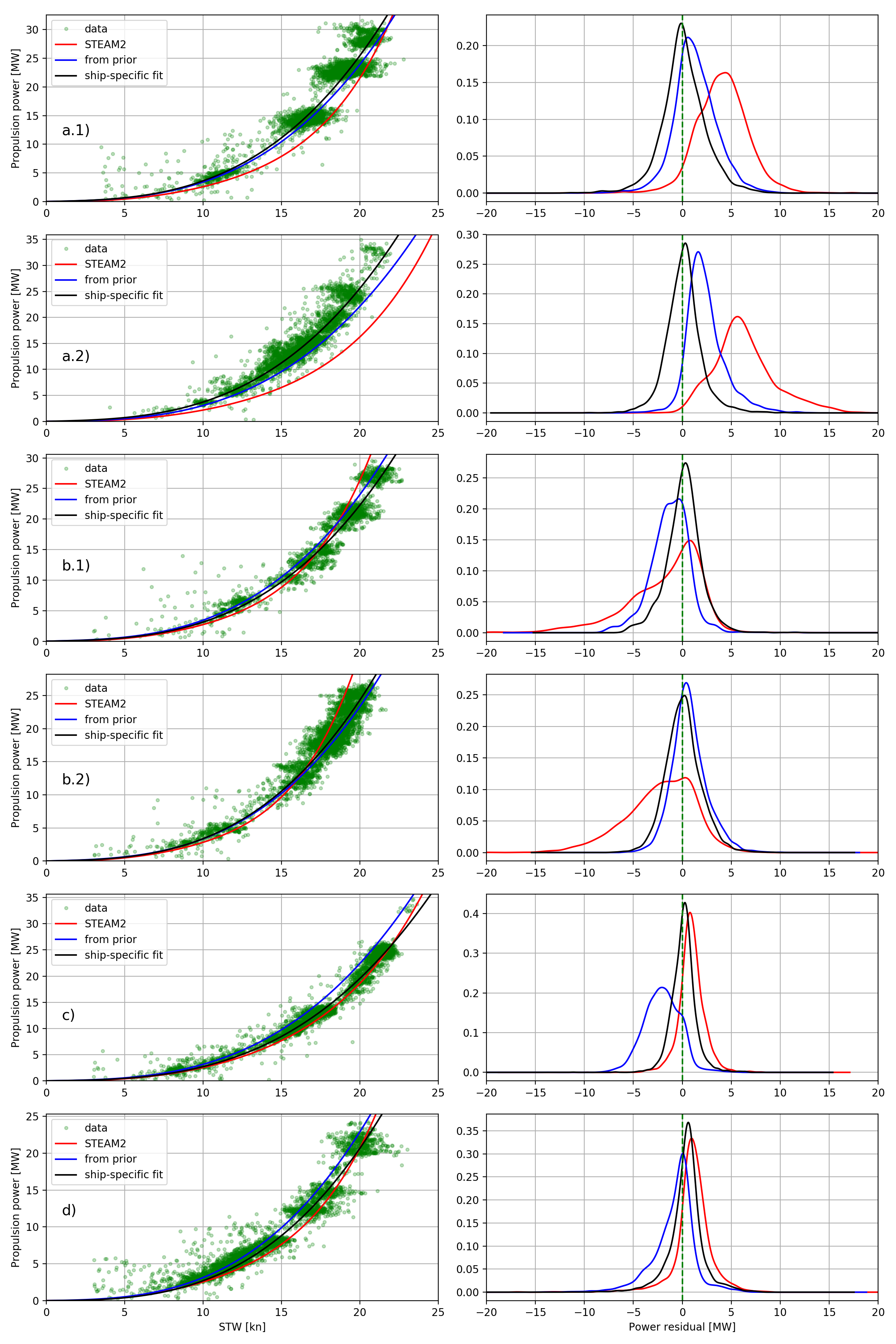}
    \caption{Left: comparison of data and speed-power curves obtained with different models. For the data-based methods, the simulation uses the median wind effect calculated from the data. Right: residual densities obtained by kernel density estimation.}
    \label{fig:speed-power}
\end{figure}

We begin by illustrating a few typical cases in \cref{fig:speed-power} by plotting the speed-power curves obtained with the different models on top of the measured data\footnote{The data-based methods also include wind as an input. Here, we simulate speed-power curves with the median of the wind effect $cos(\alpha_i)U_{R,i}^2$ calculated from the data}. In some cases, STEAM2 seems to underestimate the power consumption, e.g.\ panels a.1) and a.2) in \cref{fig:speed-power}. The prior-based model obviously has bias in several cases, but the residuals are typically smaller than for STEAM2. In some other cases STEAM2 seems to under-estimate the power with small speeds but over-estimate it with high speeds, see plots labeled with \textit{b)} in the Figure. Also in these cases the prior-based model works better in general. There are also cases where the STEAM2 model performs better or equally well than the prior-based model, see plots labeled with \textit{c)} and \textit{d)}, respectively. In all cases the ship-specific fits give the best results, which is no surprise.

\begin{figure}
    \centering
    \includegraphics[width = \textwidth]{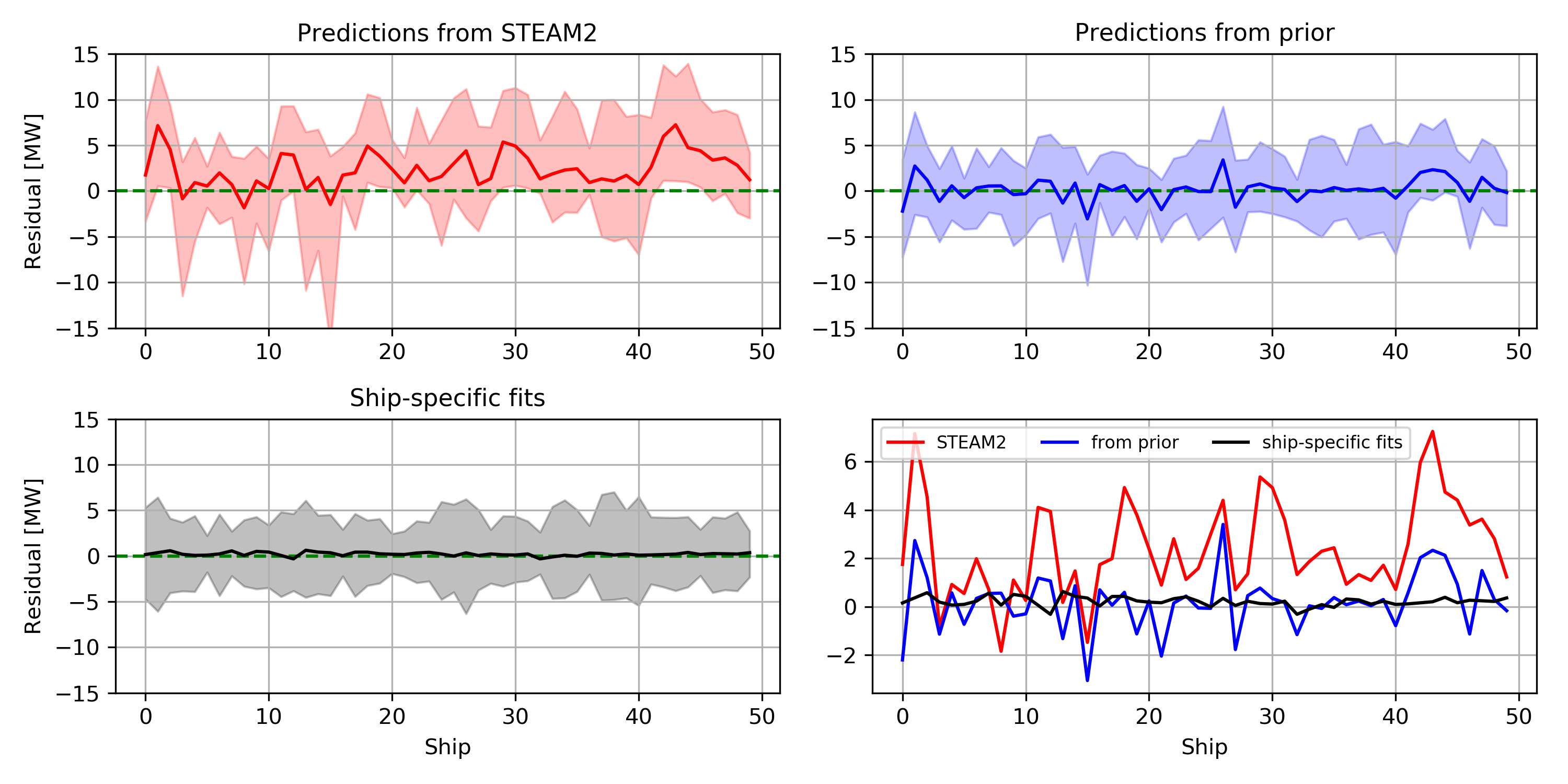}
    \caption{Residual median (solid line) and 95\% confidence region (filled area) for the different models. The bottom right figure compares the medians of the different models.}
    \label{fig:res-comp}
\end{figure}

To give a more comprehensive view on the performance of the different models, selected residual quantiles are illustrated for all the ships and all the models in \cref{fig:res-comp}. One can clearly observe the under-estimation of power in STEAM2, whereas the prior-based model residuals are more zero-centered. Thus, the hierarchical approach where the resistance coefficients of a very simple propulsion power model are predicted based only on the vessel's gross tonnage can give more accurate results than a white box approach. Note, however, that these results hold only for cruise vessels whose size is close to the range of ship sizes included in the estimation.

\begin{figure}
    \centering
    \includegraphics[width = \textwidth]{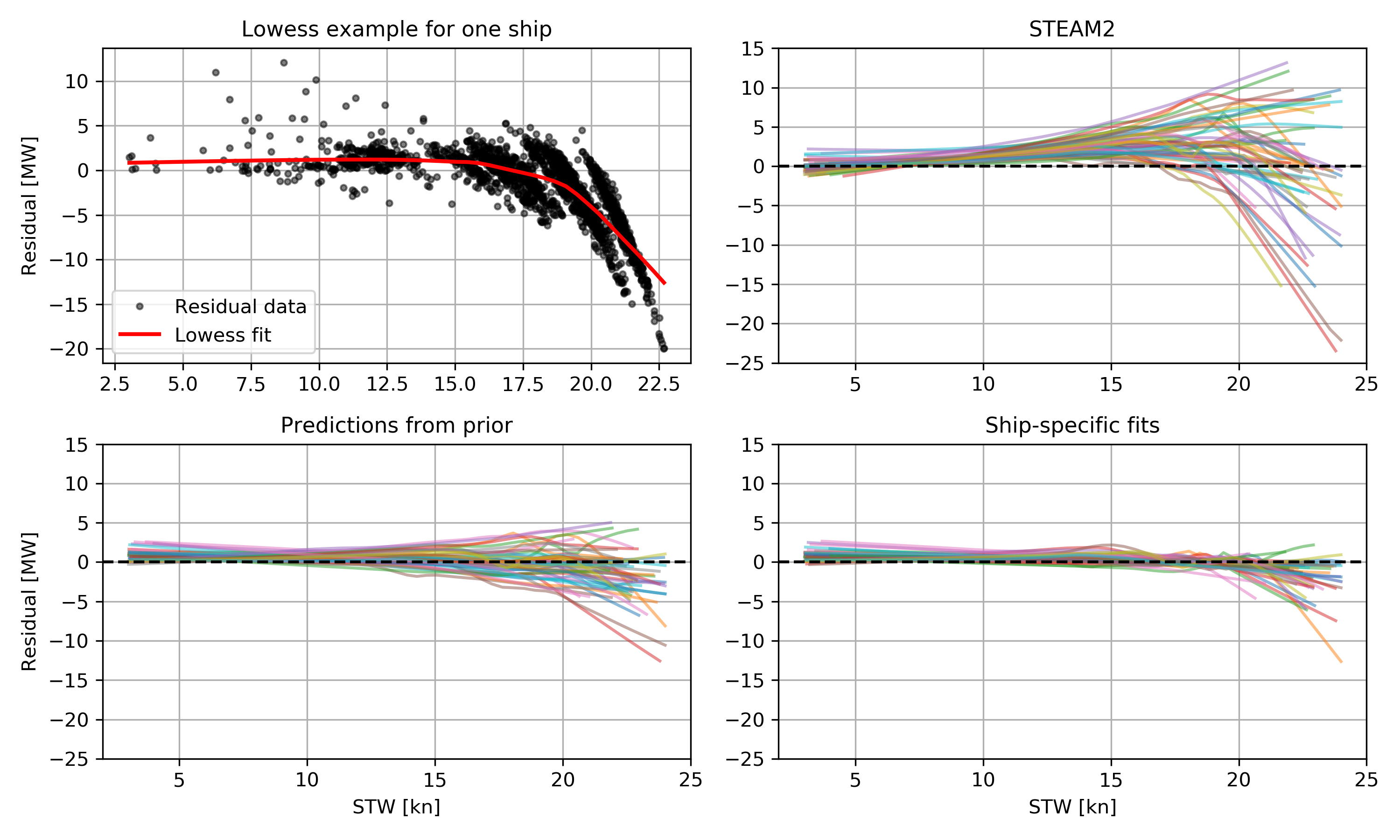}
    \caption{Residuals as a function of vessel speed. Top left: illustration of the LOWESS curve fitting to the data. Other plots: LOWESS smoothed residual vs. stw curves for different models over all ships (line color indicates a ship).}
    \label{fig:res-vs-stw}
\end{figure}

To conclude the model comparison, we illustrate in \cref{fig:res-vs-stw} how the model residuals behave as a function of vessel speed for the different models. To do this, we fit a smooth residual vs.\ speed through water (STW) curve to the data using the LOWESS method \cite{lowess} (see the top left plot in the Figure for an illustration), and then plot the smoothed curves for all the ships in one figure.  
From the Figure we can again observe the under-estimation of power in STEAM2, and also the common over-estimation of power with high speeds. Also, in line with the \cref{fig:res-comp}, the data-based models perform better and have less speed-dependent bias.

One possible factor behind the under-estimation of power in the STEAM model is that it ignores many resistance components such as wind, waves, squat and hull roughness. Work is currently underway to include many of these effects to STEAM. The data-based approach doesn't explicitly include most of these either (only wind), but since the models are fitted to the data, they calibrate to some average contribution of some of these excluded resistances. For instance, the frictional resistance coefficients in the ship-specific fits calibrate to some average hull condition over the data period included in the model fitting, and the hyper-model calibrates to some average hull condition over the ships. Moreover, wind and waves are correlated, so the wind coefficient probably ends up capturing some of the wave breaking resistance, too. In this sense, the data-based models do take these extra resistance factors into account in some way, and thus the comparison to STEAM is not completely fair. While the data-based models perform well here, within the set of cruise-ship examples, more research is needed for extrapolation to smaller vessels, where the implicitly included resistance factors may impact differently.  The purpose of this comparison is thus not to claim that the data-based methods outperform the classical white box resistance calculations, but to demonstrate that the hierarchical modeling concept provides a viable option when enough data is available.

\subsection{Obtaining statistics for predictions}

\begin{figure}
    \centering
    \includegraphics[width = \textwidth]{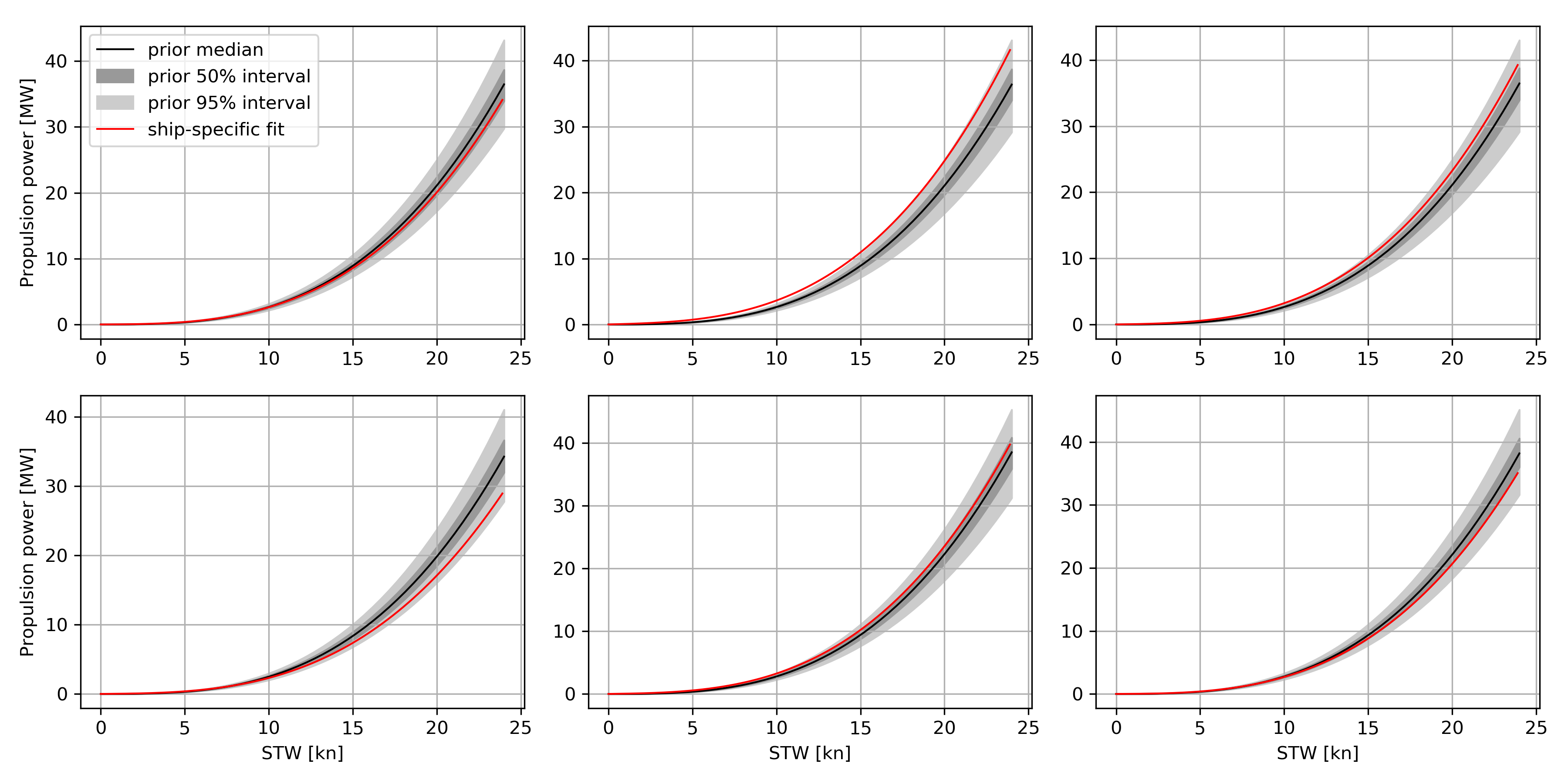}
    \caption{Confidence envelopes (50\% and 95\%) for the speed-power curves (without wind effect) predicted based in vessel's gross tonnage. The red curve comes from the ship-specific parameters and represents where the true speed-power curve roughly is.}
    \label{fig:uncertainty-demo}
\end{figure}

One benefit of the Bayesian approach is that it is statistical. Model parameters are treated as random variables, and the solution is a distribution of possible parameter values instead of point estimates. Also, this enables assessing how reliable are the estimation results and predictions made with the model. 
We demonstrate this feature by calculating the uncertainty distributions of the speed-power curves for six selected ships using the prior-based models. Due to incomplete and noisy data, there is uncertainty in the linear hyper-model parameters. Moreover, the linear model itself has errors, the magnitude of which is also estimated in the hierarchical model. Thus, with a given gross tonnage, we can give a range of values where the true speed-power curve likely is. This is illustrated in \cref{fig:uncertainty-demo}. The obtained statistics seem consistent. The "true" speed-power curve seems to fall within the calculated envelope. Wind effect was ignored here for simplicity.

\section{Conclusion and future work}
\label{sec:conclusions}

The purpose of this paper was to illustrate a hierarchical Bayesian modeling approach for marine vessels. As a prototype case, we selected cruise vessels and propulsion power prediction. For demonstration purposes, we used a simple two-parameter propulsion power model and a linear hyper-model based on vessel's gross tonnage to link together the parameters between ships. We demonstrated that the accuracy of such an approach can improve upon classical white box resistance calculation -based methods.

Calibrating these models in one go becomes computationally rather expensive when the amount of ships and data per ship increases. In practical implementations one likely needs to take another approach. One idea is to fit the models sequentially. First obtain the vessel-specific parameter estimates using the current hyper-model as the prior, and then update the hyper-parameters based on the most recent ship-specific estimates. This would give a scalable approximation to the full hierarchical model fitting.

Here we had only one "data type" in the estimations (either simulated noon-report data or high-frequency data). In real life, one would like to combine all data (both the high-fidelity and the noon report data) in the estimation. This would enable efficient borrowing of information from data rich vessels.

We feel that the results can be improved further by introducing more sophisticated propulsion power models and better hyper-models that include more ship characteristics into the estimation. In addition, we estimated only propulsion power; to get a complete picture of the vessel's fuel consumption (and thus emissions), we would need models for non-propulsion related power consumption (service power) and engine models to map power into fuel flow. An obvious topic to be analyzed in more detail is the impact of fouling effects. The main motivation of this paper was methodological, and these topics, and also generalization to other ship types, are thus left for future work.

\section*{Acknowledgments}
	
This work was supported by the Academy of Finland, decision number 313827, 'Industrial Internet and Data Analysis in Marine Industries', and by the Centre of Excellence of Inverse Modelling and Imaging (CoE), Academy of Finland, decision number 312122.  This work was  supported by the Academy of Finland, project number 334 817.This work has been supported by the European Regional Development Fund (Interreg Baltic Sea Region) project C006 CSHIPP.

\bibliographystyle{unsrt}  
\bibliography{references}  



\end{document}